# MirrorME: Implementation of an IoT based Smart Mirror through Facial Recognition and Personalized Information Recommendation Algorithm

Khandaker Mohammad Mohi Uddin[*1], Samrat Kumar Dey[†2], Gias Uddin Parvez[3], Ayesha Siddika Mukta[4], and Uzzal Kumar Acharjee[5]

[1, 2, 3, 4] Department of Computer Science and Engineering (CSE), Dhaka International University (DIU), Dhaka-1205, Bangladesh

[5] Department of Computer Science and Engineering (CSE), Jagannath University, Bangladesh

**Corresponding Author:** Khandaker Mohammad Mohi Uddin[*], and Samrat Kumar Dey[†]

**Email:** jilanicsejnu@gmail.com[*1]; sopnil.samrat@gmail.com[†2]; parvez.ka19@gmail.com[3]; siddikamukta@gmail.com[4]; and uzzal@cse.jnu.ac.bd[5]

## Abstract

We are living in the era of the fourth industrial revolution, which also treated as 4IR or Industry 4.0. Generally, 4IR considered as the mixture of robotics, artificial intelligence (AI), quantum computing, the Internet of Things (IoT) and other frontier technologies. It is obvious that nowadays a plethora of smart devices is providing services to make the daily life of human easier. However, in the morning most people around the globe use a traditional mirror while preparing themselves for daily task. The aim is to build a low-cost intelligent mirror system that can display a variety of details based on user recommendations. Therefore, in this article, Internet of Things (IoT) and AI-based smart mirror is introduced that will support the users to receive the necessary daily update of weather information, date, time, calendar, to-do list, updated news headlines, traffic updates, COVID-19 cases status and so on. Moreover, a face detection method also implemented with the smart mirror to construct the architecture more secure. Our proposed *MirrorME* application provides a success rate of nearly 87% in interacting with the features of face recognition and voice input. The mirror is capable of delivering multimedia facilities while maintaining high levels of security within the device.

## 1. Introduction

Nowadays in this world, technologies are advancing day by day. For this reason, maximum devices need to be updated with smart technology. The smart systems are organized by Artificial Intelligence (AI) and build smart equipment that makes the devices more interactive with the user. However, the smart device has the capability to easily sense, process and analyze the captured information. In modern days, working professionals are highly busy with their daily work and thus it is quite difficult for them to check daily necessary information including the latest news, To-do list, daily stock market update, social media newsfeed, traffic jam update, weather forecast and so on. This research proposed an IoT based smart mirror that helps the users to receive all this information. In addition, an AI-based face detection method also introduced that ensures a specific level of security for the proposed architecture. Besides, this system will also offer the opportunity for users to check their daily e-mails, helps them by playing preferred audio and video songs with the features of voice input. Currently, significant research works have been conducted in the domain of developing an interactive smart mirror for personal use. *Kulovic et al.* **[1]** have developed a smart mirror using the concept of Do-It-Yourself (DIY) which displays the basic information of the user including date/time, weather, and news headlines. In their development, authors have used Raspberry Pi for the development of their system and users have the privilege of adding or removing their information using their phone or tablets. Besides, by presenting the basic information of the user's, *Akshaya et al.* **[2]** have developed a two-button supported smart mirror, one for the website and another for the map. The key feature of this research is to access the smart mirror from several areas with an assistance of a user-friendly interface. In other work, *Yusri et al*. **[3]** introduced a smart mirror that offers users to control their home appliances using a voice recognition mechanism. Mainly, this application is developed to save the energy consumptions of the users along with showing some basic features including time, traffic, weather, and so on. Specifically for the aged and disabled peoples, *Athira et al.* **[4]** have developed a mirror using passive Infrared (PIR) sensors. The designed system highly capable of detecting a person with the support of internet connections. Moreover, basic features including social media notifications, weather information, and time also incorporated with this architecture. In other work, in order to present the basic features and to control the household devices, *Hossain et al.* **[5]** design a smart mirror. However, to verify an authenticated user that article discussed the mechanism of using face recognition technology. Similarly, *Nadaf et al*. **[6]** and *Njaka et al.* **[7]** also developed a smart mirror for ensuring home security along with some basic features. A biometric authentication based user recognition model has developed by the authors **[7-8]** to ensure sufficient security in accessing the smart mirror. To detect an intruder through a smart mirror in a room, *Jin et al.* **[8]** have proposed an alarm based interactive application. This article focuses on storing images of users in a dataset and the designed system send notifications to the authenticated user when the model does not identify someone.

With an 80% accuracy, in detecting users faces *Mohamed et al.* **[9]** have designed an artificial intelligence-based smart mirror. For recognizing existing users, they also proposed a registration method with the support of the facial recognition method. In recent time, *Hollen et al.* **[10]** have designed a facial recognition-based smart mirror to detect the user mood in real-time. The authors employed the concept of the user's face and outbound movement's detection to recognize the face and identify the mood in the system. In this article, design and implementation of an IoT based smart mirror with the support of personalized information recommendation and face recognition approach is proposed. Following **Table 1** highlights the available features for both category of users (general users and authenticated users). This research have focused on the design and development of an interactive smart mirror; development of a voice-controlled input; and to ensure the accurate detection of users faces to access the features.

**Table 1:** User wise features availability of *MirrorME* architecture

| | General Users | Authenticated Users |
|---|---|---|
| Features Category | Time, weather, calendar, alarm, news update, COVID-19 update, YouTube, music, and traffic update. | Gmail, stock market update, To-do list, phone notification, YouTube channel, and all the available features of the general user. |

## 2. Methodology

In this section, the design methodology of this proposed framework has been discussed. The proposed architecture is suitable for the essential functionality of any smart home. On the other hand, this will timely remind the user for a specific task based on the user's preference. Different features of the proposed *MirrorME* application are illustrated in **Figure 1**. *MirrorME* is designed dedicatedly for two different categories of users; one is general users and other is authenticated users. The authentication process is developed with the support of HOG-SVM based face recognition approach. Regarding the access of different features, there is some personal information recommendation features for any existing users whereas the general users only have the access of some specific basic features. However, the authenticated users also access some of the premium features of smart mirror by using the voice recognition features. Functional and access diagram of general user and authenticated user of *MirrorME* application is shown in **Figure 2**.

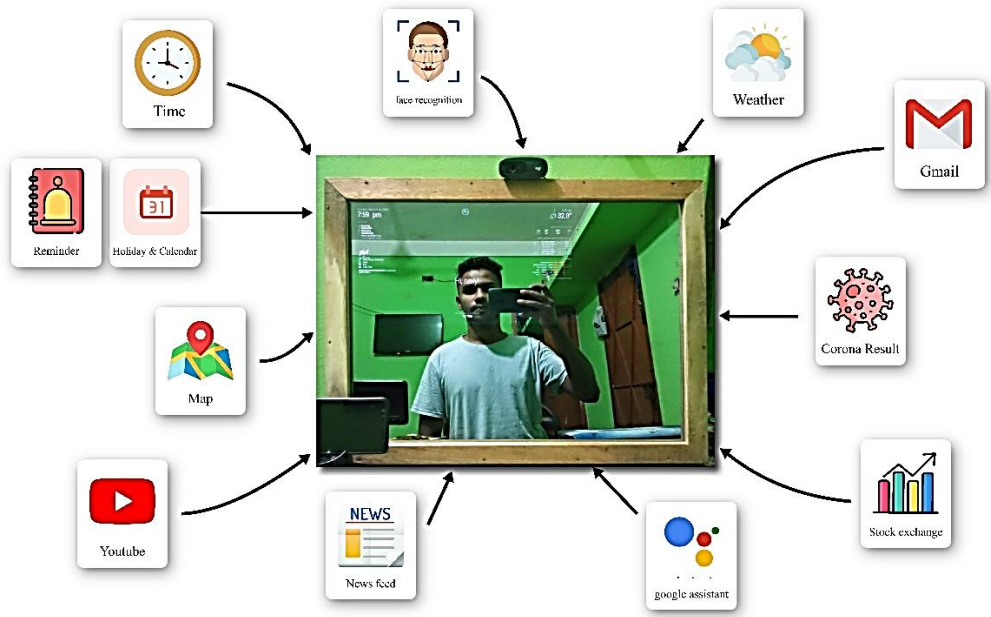

**Figure 1.** Various features of the proposed *MirrorME* application

As we have already discussed, in this exploration, face recognition is required to detect the authentic user for accessing the smart mirror. Face recognition has been employed in this research using the architecture of Histograms of Oriented Gradients (HOG) features and the linear Support Vector Machine (SVM). The proposed feature extraction algorithm by *Dalal et al.* **[11]** known as Histograms of Oriented Gradients (HOG). In HOG, initially the image is divided into small cells and then the gradient or edge direction histogram is collected in each cell unite, and finally these histograms are combined to form a HOG descriptor. While doing feature extraction, HOG generally creates 3780 features of the image.

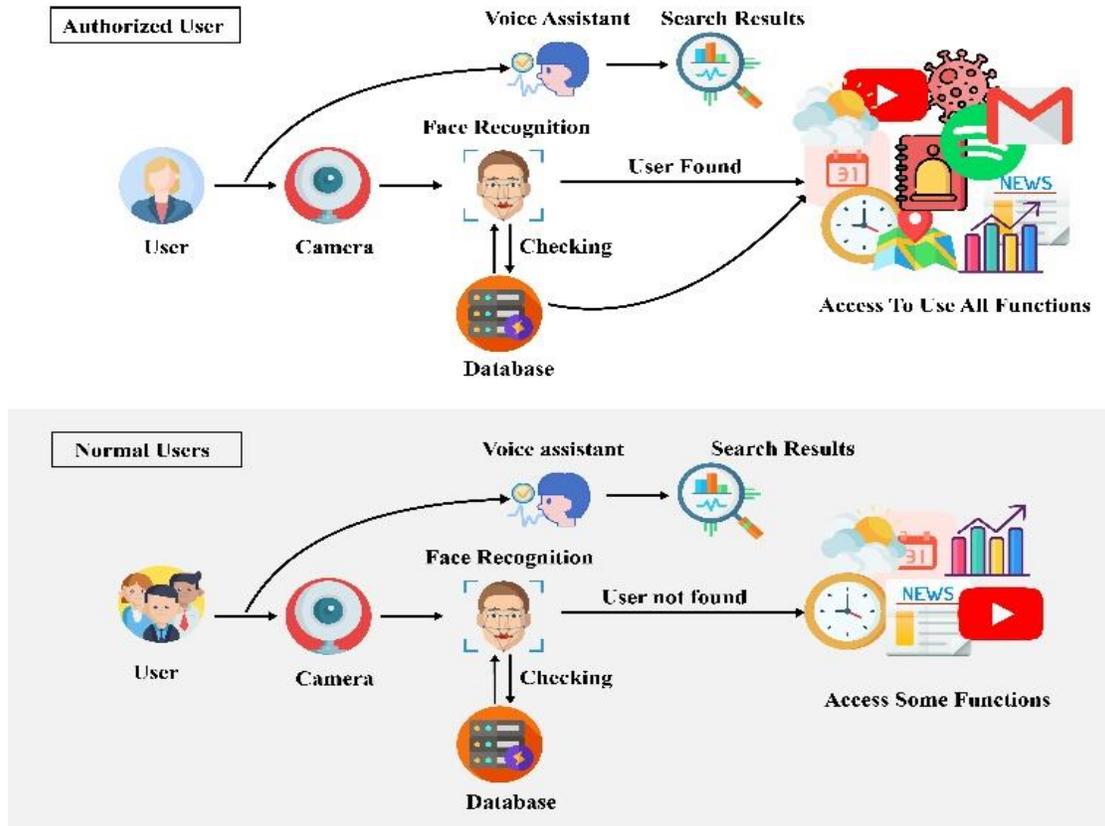

**Figure 2.** Functional and access diagram of general user and authenticated user of *MirrorME* application

A gradient in the horizontal and vertical directions is obtained first to extract the feature in HOG. The template of horizontal direction is $K = [-1,0,1]$ and to filter image, its transposition can be used, that is why it is possible to calculate the horizontal and vertical gradients, easily. The expression **(Eq. 1)** and **(Eq. 2)** is given as follows

$$g_x = I(x+1, y) - I(x-1, y) \tag{1}$$
$$g_y = I(x, y+1) - I(x, y-1) \tag{2}$$

Here, the direction gradient of x is represented by $g_x$ and the direction gradient of y is represented by $g_y$, the pixel value of $(x, y)$ is denoted by $I(x, y)$. The gradient magnitude of $(x, y)$ is denoted by $\Delta g\ (x, y)$ and calculated **(Eq. 3)** by

$$\Delta g\,(x,y) = \sqrt{g_x{}^2 + g_y{}^2} \tag{3}$$

And, the gradient direction (ϴ) of $(x, y)$ is calculated **(Eq. 4)** by

$$\theta = \arctan(g_y/g_x) \tag{4}$$

In machine learning, Support Vector Machine (SVM) **[12]** used as a binary classifier. Classes separated with the largest gap between the support vectors. Here, the borderline instances in a class known as support vectors. In addition, we have extended the SVM with the support of kernel. Kernels transform data from input space to feature space. The kernel is a mathematical function that takes two arguments and returns the dot product of their value. Let two data points $x_1$ and $x_2$ mapping is denoted by $\varphi$ then the kernel $K$ will be **(Eq. 5)**

$$K(x_1, x_2) = \varphi(x_1)\,T\,\varphi(x_2) \tag{5}$$

When input space and feature space are equal then it is known as linear kernel and this linear kernel is used in linear SVM. Mathematically, the equation of linear kernel can be expressed as **(Eq.6)**

$$K(x_1, x_2) = x_1\,T\,x_2 \Rightarrow \varphi(x) = x \tag{6}$$

Due to its capability of faster application training therefore, it is highly efficient in high-dimensional data applications. The proposed workflow is consisted of detecting faces using HOG and linear SVM, computing embedding and comparing the vector to the database via a voting method. The basic flow of the HOG feature extraction algorithm and the face detection technique of the proposed application is illustrated in **Figure 3.**

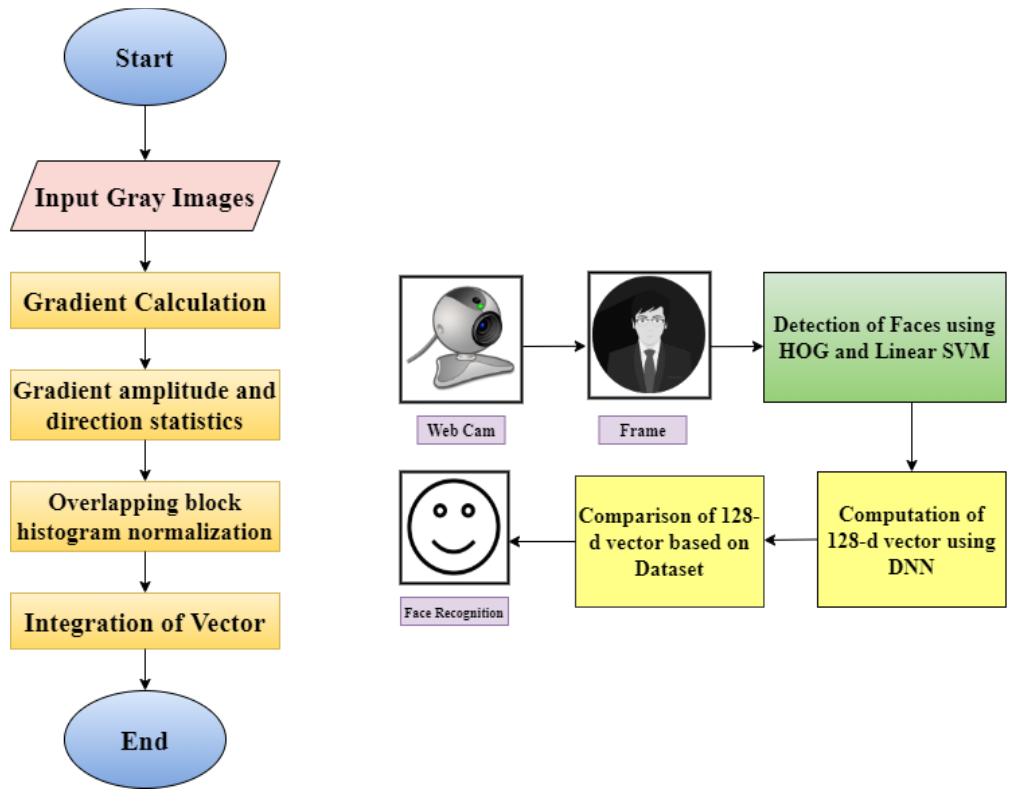

**Figure 3.** Workflow diagram of feature extraction algorithm and face detection procedure using Histogram of oriented gradients (HOG) and Support Vector Machine (SVM)

## 3. Design and Implementation

In this segment, the design and implementation approach has discussed with the architectural view of hardware and software. The hardware of the proposed framework consists of a various components including two-way mirror, raspberry pi 3B+, camera, microphone, wood frame, display screen, speaker and some others. Following **Figure 4** depicts the IoT based hardware configuration diagram of the proposed *MirrorME* application.

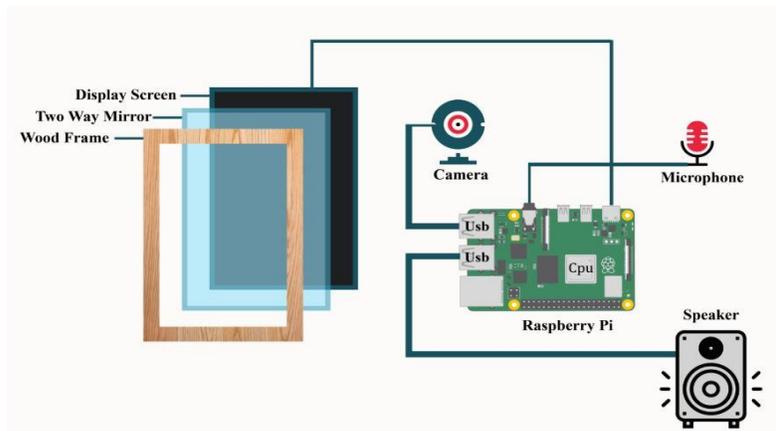

**Figure 4.** IoT based hardware configuration diagram of the proposed *MirrorME* application

### 3.1 Raspberry pi

The Raspberry pi [13] is a small and powerful computer that uses its own OS known as Raspberry pi OS and this operating system is Debian-based. Importantly, this device is the core component of our designed smart mirror.

### 3.2 Camera

For the purpose of face recognition, this work needed to use a camera and our proposed algorithm is incorporated with the camera to identify an existing user for accessing the *MirrorME* application.

### 3.3 Two way mirror

Basically, a mirror is a smooth polished surface where the image is created by the refection. In this work, a two-way mirror is used in which one side is transparent and the other side shows its reflection. The concept of using a two-way mirror is similar to using a usual mirror and that same mirror also acts as a functional object.

### 3.4 Display monitor

In order to present the basic information including date/time, weather update, alarm, news headlines, traffic update and so on, a display monitor is essential in this model. Therefore, in the development of the model, LG 14 inch monitor is connected to the Raspberry pi module using HDMA interfacing.

## 3.5 Microphone and speaker

To utilize the features of the voice-based activity, a microphone is needed in the designed model. A single Bluetooth speaker is also attached to the system to receive feedback. However, all the output results can be received through the speaker.

Based on the above discussion, the complete model of the architectural diagram and framework design approach of the proposed system shown in **Figure 5.**

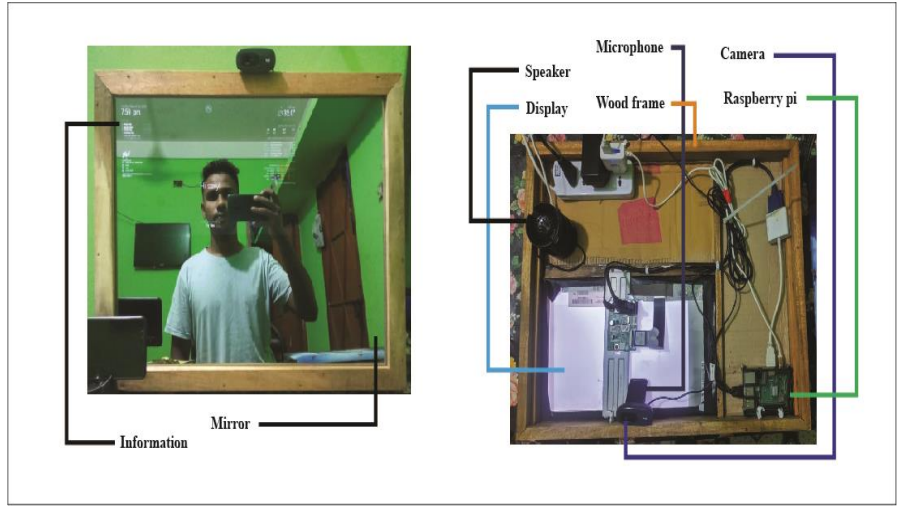

**Figure 5**. Architectural diagram and framework design approach of the proposed system

To avoid unnecessary access to personal features for specific users, this model has also designed a multi-phase user recognition system. For recognizing the valid and existing users, HOG (Histogram of Oriented Gradients) and Linear SVM (Support Vector Machine) based face recognition model have been employed. The detailed working procedure of the proposed system highlighted in **Figure 6.** In addition, a stepwise description of the working procedure presented in **Table 1.**

**Table 1:** Working procedure of *MirrorME* application based on user activity

| Order | Task | Description |
|---|---|---|
| STEP 1 | Performing the face detection | Initially, the designed system will activate based on internet connectivity. It will continuously check for an internet connection to perform the face detection and to check whether the user is existing user or general user. |
| STEP 2 | User authentication check | Through this step, designed systems will be able to distinguish between general user and existing user. |

| STEP 3 | Showing basic features for general user using voice input | For a new user, this application will only offer some of the basic features including date/time, weather information, calendar, holiday and other information. |
|---|---|---|
| STEP 4 | Check existing users through face recognition and voice input | In this step, after successful detection of existing user, designed application will provide access to all the premium features. |
| STEP 5 | Authenticated users has the privilege to access all the features of the application | Besides, for the authenticated users, the system will also provide the user interaction features to check their regular email, stock exchange update, and SMS notification. |

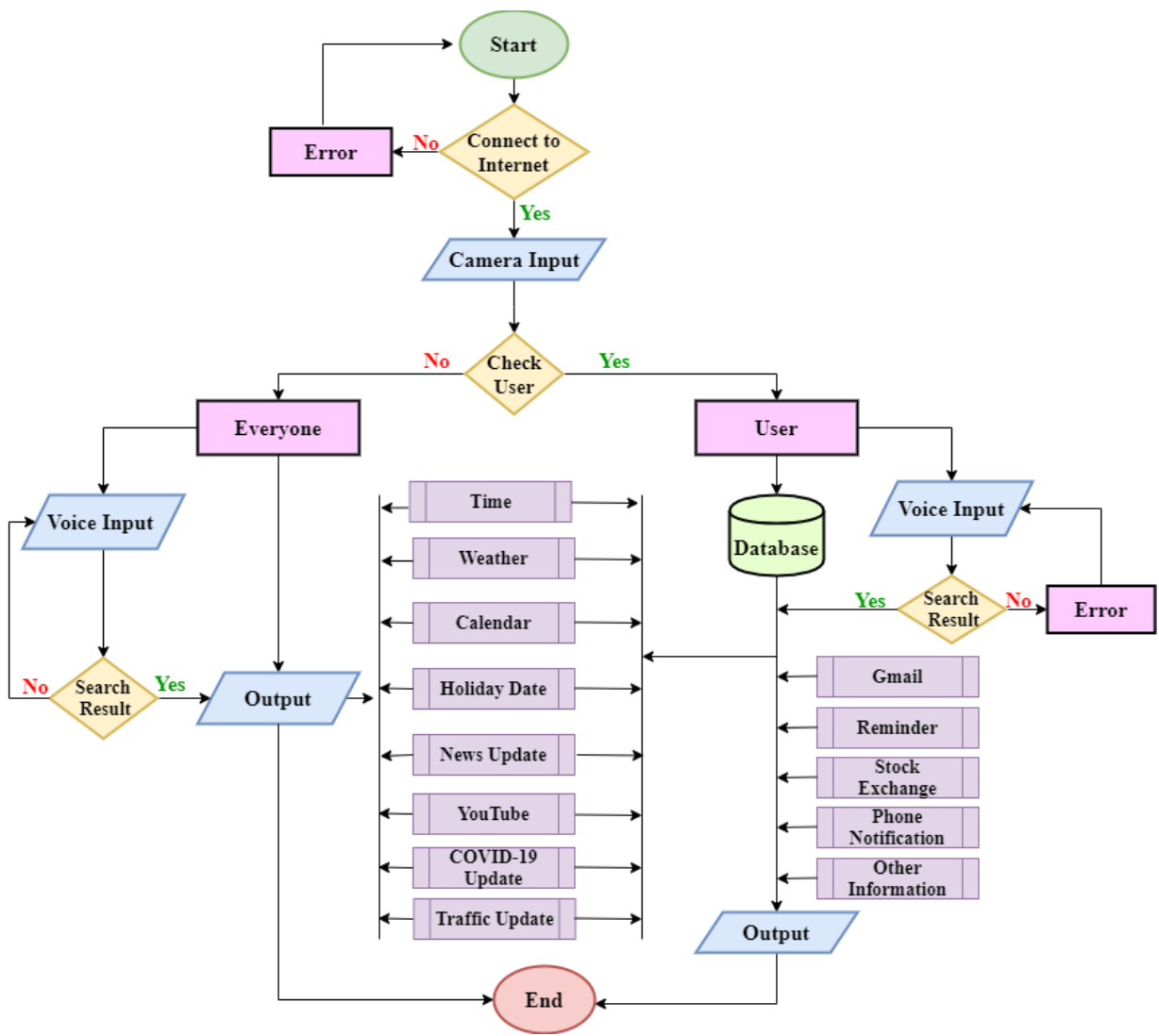

**Figure 6.** A complete flowchart of *MirrorME* application working procedure with the features of different user interactions

Following **Figure 7** illustrates both the software view and implantation process of *MirrorME* application. In Figure 7(A), all the available activity is properly labelled with appropriate features. Apart from designing the hardware architecture, this research also developed a dedicated software view for a better understanding of this application. On the other hand, in the implemented view 7(B), 7(C) and 7(D) depicts the process of using home interface, road traffic update, and YouTube videos respectively.

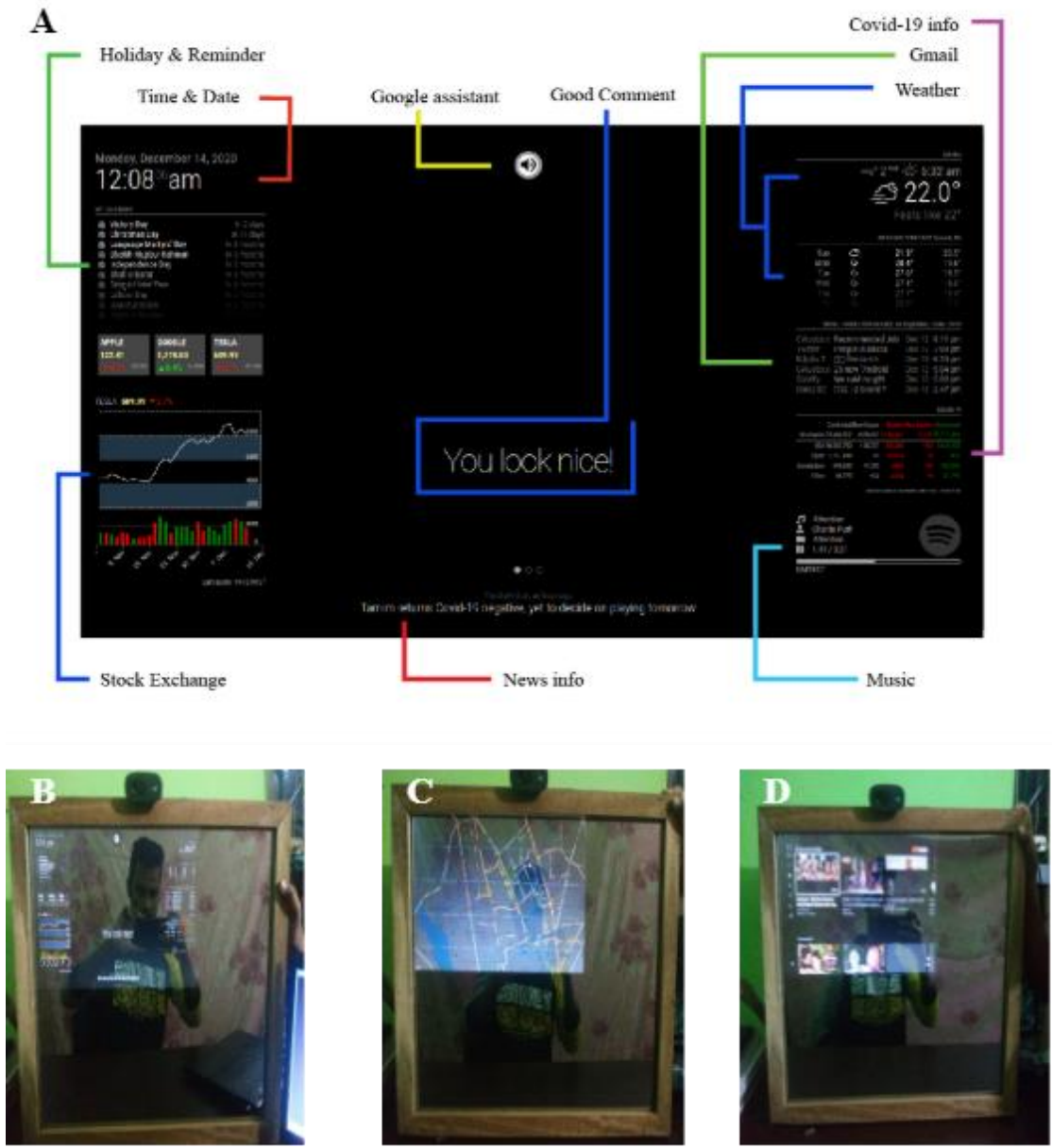

**Figure 7.** Software view and implemented view of *MirrorME* application

## 4. Results and Discussion

This section has discussed testing results based on experiments. In order to test our designed model, we have conducted a result analysis based on the experience users observed during the assessment of this application. This research work has been carried out in the machine intelligence lab (MINTEL) of the Dhaka International University. Ten (10) participants have chosen to participate in this experiment based on previous experience in interacting with smart devices. All the participants were chosen aged between 22-25 years. In order to achieve higher accuracy in the task of face recognition, we have chosen a frame rate of 50 and a resolution of 480 x 360. The resolution 480×360 has been taken because a high-resolution frame takes a higher time to process and a low-resolution frame normally does not provide high accuracy. Among the Ten (10) percipients, there were Four (4) female participants (FP) and Six (6) male participants (MP) to interact with the *MirrorME* system. For each participant, exactly Ten (10) responses have been recorded for the features of face recognition and others. The success rate for each individual is shown in **Table 2.** However, during the testing phase, face recognitions features shows ~100% accuracy for the fourth male participants (MP4). On the other hand, it shows an accuracy of ~60% for the first female participants (FP1).

**Table 2:** Analyzing the accuracy/success rate of face recognition feature based on the participations of (n=10) individuals

| Face recognition | Number of terms | Success rate |
|---|---|---|
| **Response form Male Participant** | | |
| MP1 | 10 | 90% |
| MP2 | 10 | 70% |
| MP3 | 10 | 80% |
| MP4 | 10 | 100% |
| MP5 | 10 | 90% |
| MP6 | 10 | 80% |
| **Response form Female Participant** | | |
| FP1 | 10 | 60% |
| FP2 | 10 | 70% |
| FP3 | 10 | 80% |
| FP4 | 10 | 90% |

After the testing phase of face recognition, the same ten (10) participants again interacted with the mirror with the voice input. Each of the participants provides ten (10) voice input to interact with the individual features including YouTube, Alarm, Traffic, and daily schedule. **Table 3** presents the testing results of the voice input. However, due to the network connectivity issue, sometimes the voice input took a longer time to function properly.

Table 3: Analyzing the accuracy/success rate of voice input feature based on the participations of (n=10) individuals

| Participants for Voice command | Number of terms | Success rate: YouTube | Success rate: Alarm | Success rate: Traffic | Success rate: Daily schedule update |
|---|---|---|---|---|---|
| **Response form Male Participant** | | | | | |
| MP1 | 10 | 80% | 90% | 80% | 90% |
| MP2 | 10 | 90% | 100% | 100% | 100% |
| MP3 | 10 | 80% | 90% | 90% | 80% |
| MP4 | 10 | 90% | 100% | 90% | 90% |
| MP5 | 10 | 70% | 80% | 80% | 90% |
| MP6 | 10 | 80% | 90% | 90% | 90% |
| **Response form Female Participant** | | | | | |
| FP1 | 10 | 70% | 80% | 80% | 90% |
| FP2 | 10 | 80% | 90% | 90% | 80% |
| FP3 | 10 | 90% | 90% | 100% | 80% |
| FP4 | 10 | 80% | 90% | 80% | 90% |

For each participant, we have recorded exactly (10×10) responses for both face recognition and voice input features. After the result analysis, we observed the average success rate of ~86.75% during the time of interaction with the *MirrorME*. **Table 4** shows the average success rate for each individual based on the data of **Table 2** and **Table 3**. In addition, we also highlighted the three highest average success rate for the second male participant (~97%), fourth male participant (~92%) and third female participant (~90%).

Table 4: Average success rate of each individual participants for both the face recognition and voice input

| Participants | Number of terms | Average success rate |
|---|---|---|
| MP1 | 10 | 85% |
| MP2 | 10 | 97.5% |
| MP3 | 10 | 85% |
| MP4 | 10 | 92.5% |
| MP5 | 10 | 80% |
| MP6 | 10 | 87.5% |
| FP1 | 10 | 80% |
| FP2 | 10 | 85% |
| FP3 | 10 | 90% |
| FP4 | 10 | 85% |
| FP1 | 10 | 85% |

A comparative analysis section also designed and shown in **Table 5** to compare the features of *MirrorME* application with other research work. Based on the comparison, we observed that our proposed system contains every feature that considered as a complete interactive smart mirror and has outperformed other work in the specific category.

**Table 5:** A comparative features analysis of designed *MirrorME* application with other research work

| Article /Features | Basic Information | Reminders | Social Media Notifications | Voice Recognition | Voice Control | Face Recognition | Email Notifications | Google Map | Weather Update |
|---|---|---|---|---|---|---|---|---|---|
| Kulovic et al. [1] | √ | | | | | | | | |
| Akshaya et al.[2] | √ | | | | | | | √ | |
| Yusri et al. [3] | √ | | | √ | | | | | |
| Athira et al. [4] | √ | √ | √ | √ | | | | | |
| Hossain et al.[5] and Nadaf et al.[6] | √ | | | | | √ | | | |
| | √ | | | | √ | √ | | | |
| Njaka et al. [7] and Jin et al. [8] | √ | | | | | √ | √ | | |
| Mohamed et al. [9] | √ | | | | | √ | √ | √ | |
| Hollen et al. [10] | | | | | | √ | | | √ |
| Our Proposed System (MirrorME) | √ | √ | √ | √ | √ | √ | √ | √ | √ |

## 5. Conclusion

This paper demonstrated a smart mirror device with a user-friendly architecture with many impressive features. Following a service-oriented approach, a stable and easy-to-use architecture was also introduced in this article. However, security issues cannot be overlooked in today's world of interconnected devices. Therefore, this system was designed with a strong authentication framework to ensure the system's end-to-end security. Two different features (face recognition and voice input) sets this designed model aside from other similar works. Furthermore, unique features such as information customization for each individual user made this model more convenient and effective. Again, the prototype's advancement has infinite possibilities in the future in the field of medical data analysis including BMI calculations, temperature check and blood pressure indicator. The most notable aspect of this model is the ability to carry the smart mirror display across the entire house.